\documentclass[11pt]{article}
\usepackage[T1]{fontenc}
\usepackage[utf8]{inputenc}
\usepackage[english]{babel}

\usepackage{hyperref}
\usepackage{geometry} 
\usepackage{authblk}

\usepackage{amsmath,amssymb,amsfonts}
\usepackage{amsthm,bbm}
\usepackage{braket}

\usepackage{graphicx}
\usepackage{enumerate}
\usepackage{algorithm}
\usepackage{algpseudocode}
\usepackage{booktabs}
\usepackage{accents}

\usepackage{cite}

\usepackage[font=small,labelfont=bf]{caption}
\newcommand{\lbar}[1]{\underaccent{\bar}{#1}}
\newcommand{\rank}[0]{\mathrm{rank}}
\newcommand{\ie}[0]{\textit{i.e. }}
\newcommand{\eg}[0]{\textit{e.g. }}
\newcommand{\Pset}[0]{\Delta_d^\downarrow}
\newcommand{\maj}[0]{\succeq}
\newcommand{\nmaj}[0]{\nsucceq}
\newcommand{\free}[0]{\underset{\mathrm{free}}{\rightarrow}}
\newcommand{\bifree}[0]{\underset{\mathrm{free}}{\leftrightarrow}}
\newcommand{\nbifree}[0]{\underset{\mathrm{free}}{\nleftrightarrow}}
\newcommand{\Free}[0]{\mathcal{F}}
\newcommand{\E}[0]{\mathcal{E}}
\newcommand{\T}[0]{\mathcal{T}}
\renewcommand{\P}[0]{\mathcal{P}}
\renewcommand{\S}[0]{\mathcal{S}}
\newcommand{\ocr}[0]{\mathrm{ocr}}
\newcommand{\La}[0]{\mathcal{L}}
\newcommand{\K}[0]{\mathcal{K}}
\newcommand{\up}[0]{\mathrm{up}}

\theoremstyle{definition}
\newtheorem{definition}{Definition}
\newtheorem{lemma}{Lemma}
\newtheorem{proposition}{Proposition}
\title{Optimal common resource in majorization-based resource theories}

\author[1]{G M Bosyk\thanks{gbosyk@fisica.unlp.edu.ar}}
\author[2]{G Bellomo}
\author[1]{F Holik}
\author[3]{H Freytes}
\author[3]{G Sergioli}

\affil[1]{\small Instituto de F\'{\i}sica La Plata, UNLP, CONICET, Facultad de Ciencias Exactas, 1900 La Plata, Argentina}
\affil[2]{\small CONICET-Universidad de Buenos Aires, Instituto de Investigación en Ciencias de la Computación (ICC), Buenos
Aires, Argentina}
\affil[3]{\small Universit\`{a} degli Studi di Cagliari, Via Is Mirrionis 1, I-09123, Cagliari, Italy}

\date{}

\begin{document}

\maketitle

\renewcommand{\abstractname}{\vspace{-\baselineskip}}

\vspace{-13mm}
\begin{abstract}
We address the problem of finding the optimal common resource for an arbitrary family of target states in quantum resource theories based on majorization, that is, theories whose conversion law between resources is determined by a majorization relationship, such as it happens with entanglement, coherence or purity.
We provide a conclusive answer to this problem by appealing to the completeness property of the majorization lattice.
We give a proof of this property that relies heavily on the more geometric construction provided by the Lorenz curves, which allows to explicitly obtain the corresponding infimum and supremum.
Our framework includes the case of possibly non-denumerable sets of target states (i.e.,
targets sets described by continuous parameters). In addition, we show that a notion of approximate majorization, which has recently found application in quantum thermodynamics, is in close relation with the completeness of this lattice.
Finally, we provide some examples of optimal common resources within the resource theory of quantum coherence.

\end{abstract}




\section{Introduction}

Quantum resource theories (QRTs) are a very general and powerful
framework for studying different phenomena in quantum theory from an
operational point of view (see Ref.~\cite{Chitambar2019} for a
recent review of the topic). Indeed, all QRTs are built from three
basics components: free states, free operations and resources. These
components are not independent among each other, and they are
defined in a way that depends on the physical properties that one
wants to describe. In general, for a given QRT, one defines the set
of free sates $\Free$, formed by those states that can be generated
without too much effort. Then, an operation $\E$ is said to be
\emph{free}, if it satisfies the condition of mapping free states
into free states: $\E$ is free if and only if
$\E(\rho) \in \Free \ \forall \ \rho \in \Free$. Thus, free
operations can be interpreted as the ones that are easy to implement
in the lab. Finally, quantum resources are defined as those states
that do not belong to the set of free states (i.e., $\rho \notin
\Free$). These states are the useful ones for doing the
corresponding quantum tasks. As an illustration, consider the task of transmitting an arbitrary quantum state from one lab to another distant one, where the allowed free operations are the so-called local operations and classical communication  (LOCC).
In this typical scenario, entanglement arises as the necessary quantum resource to perform this task (as it can be seen from the quantum teleportation protocol~\cite{Bennett1993}).

Clearly, it is not possible to convert free states into resources by
appealing to free operations alone. This is the reason why the term
\emph{resource theory} was coined. In fact, one of the main concerns
of the QRTs is the characterization of transformations between
resources by means of free operations. Here, we are focused on QRTs
for which these transformations are fully characterized by a kind of
\textit{majorization law} between the resources. Precisely, we are
interested in QRTs for which $\rho \free \sigma$ is equivalent to
$x(\rho) \maj x(\sigma)$ or $x(\sigma) \maj x(\rho)$, where
$x(\rho)$ and $x(\sigma)$ are probability vectors associated to
$\rho$ and $\sigma$, respectively, and $\maj$ means a majorization
relation (see \eg~\cite{MarshallBook} for an introduction to
majorization theory). In addition to the characterization of the
convertibility of free states by means of free
operations~\cite{Nielsen1999,Du2015,Chitambar2016,Du2017,Zhu2017,Gour2015,Streltsov2018},
majorization theory has been applied to different problems in
quantum information such as entanglement
criteria~\cite{Nielsen2001a,Partovi2012}, majorization uncertainty
relations~\cite{Friedlan2013,Puchala2013,Rudnicki2014,Luis2016,Rastegin2016},
quantum entropies~\cite{Wehrl1978,Bosyk2016,Hanson2018} and quantum
algorithms~\cite{Latorre2002}, among
others~\cite{Nielsen2000,Nielsen2001b,Nielsen2001c,Chefles2002,Bellomo2019}.

We restrict to QRTs based on majorization mainly for two reasons. As
we have already mentioned, there are several examples of QRTs that
satisfy a majorization law (see Table~\ref{tab:QRT maj} and
Refs.~\cite{Nielsen1999,Du2015,Chitambar2016,Du2017,Zhu2017,Gour2015,Streltsov2018}).
Thus, the results obtained which are based in the properties of
majorization are of great generality, providing a unifying framework
for several physical problems. On the other hand, majorization
induces a \emph{lattice
structure}~\cite{Bapat1991,Bondar1994,Cicalese2002}. We will show
that this allows to introduce the notion of \emph{optimal common
resource} in a very natural way. Before doing that, we stress that
the lattice theoretical aspects of majorization theory have not been
sufficiently exploited in comparison with other features of it in
the area of quantum information. Indeed, the first applications were
given in Refs.~\cite{Partovi2009,Partovi2011}; only recently, new
applications of the majorization lattice have been found~\cite{Bosyk2017,Korzekwa2017,Bosyk2018,Sauerwein2018,Wang2018,Guo2018,Yu2019,Li2019}.

\begin{table}
\centering
\resizebox{\textwidth}{!}{%
\begin{tabular}{@{}lllll@{}}
\toprule
\multicolumn{1}{c}{\textbf{QRT}}  & \multicolumn{1}{c}{\textbf{Free operations}} & \multicolumn{1}{c}{\textbf{Resources}} & \multicolumn{1}{c}{\textbf{Probability vector}} \\ \midrule
Entanglement (pure) \cite{Nielsen1999} & LOCC & $\ket{\psi} = \sum_i \sqrt{\psi_i} \ket{i^A}\ket{i^B} \in \mathbb{C}^{d_A} \otimes \mathbb{C}^{d_B}$ (Schmidt decomposition) &
$x(\psi) \equiv [\psi_1, \ldots, \psi_d]$ with $d = \min\{d_A,d_B\}$ \\
Coherence (pure) \cite{Du2015,Chitambar2016,Du2017,Zhu2017}& IO & $\ket{\psi} = \sum_i \psi_i \ket{i} \in \mathbb{C}^{d} $ ($\{\ket{i} \}$ incoherent basis) &
$x(\psi) \equiv [|\psi|^2_1, \ldots, |\psi|^2_d]$
\\
Purity \cite{Gour2015,Streltsov2018} & Unital & $\rho \neq \frac{I}{d}$ acting on $\mathbb{C}^{d}$ &
 $x(\rho) \equiv [\rho_1, \ldots, \rho_d]$ with $\rho_i$ eigenvalues of $\rho$\\
\bottomrule
\end{tabular}%
}
\caption{Quantum resource theories where the transformations between resources by means of free operations are given by a majorization relation.
For each QRT, the corresponding free operations are: local operations and classical communication (LOCC), incoherent operations (IO) and unital, respectively.}
\label{tab:QRT maj}
\end{table}

Here, we aim to address the following problem. Let us suppose that
one wants to have a set of target resources $\T$. For obvious practical
reasons, it is very useful to find a resource $\rho$, such that it
can be converted by means of free operations to any other resource
belonging to the target set, that is, $\rho \free \sigma$ for all
$\sigma \in \T$. By definition, the \emph{maximal resource} (if it
exists) has to perform this task for any target set in a given QRT. But a more interesting question is whether there exists
a state that can carry out the same task, but using the least amount
of resources as possible. More precisely, one aims to find a resource
$\rho^{\ocr}$ such that $\rho^{\ocr} \free \sigma \ \forall \sigma
\in \T$, and for any other $\rho$ satisfying $\rho \free
\sigma \ \forall \sigma \in \T$, then either $\rho \free \rho^\ocr$
or $\rho \nbifree \rho^\ocr$.
If this state exists, we refer to it
as the \textit{optimal common resource} (ocr). \emph{In this work,
we provide a solution for the problem of finding the optimal common
resources for arbitrary target sets of all QRTs based on
majorization}.
This problem was already posed and (partially) solved in
Ref.~\cite{Guo2018}, for possibly infinite (but denumerable) target
sets of bipartite pure entangled states. Let us stress that our
proposal is a twofold extension of that previous work. In the first
place, we provide a unifying framework for arbitrary QRTs based on
majorization, which includes not only entanglement resource theory,
but also the important cases of coherence and purity resource
theories. In the second place, we consider the most
general case of possibly non-denumerable sets of target resources.
This is a powerful extension of previous works, because it allows to apply this technique to target sets which are described by a continuous family
of parameters.
We provide the answer to this general problem by appealing to the completeness of the majorization
lattice \cite{Bapat1991,Bondar1994}.
In particular, our construction relies on the geometrical properties of Lorenz curves associated to the corresponding target set of probability vectors, which allow us to provide an explicit algorithm for the computation not only of the infimum (as in \cite{Bapat1991}) but also of the supremum.
We also describe, for convex polytopes, the relationship between the
infimum and supremum and their extreme points.

\section{Majorization lattice}

Here, we introduce the majorization lattice and
present its most salient order-theoretic features.

Let us consider probability vectors whose entries are sorted in
non-increasing order, that is, vectors belonging to the set:
\begin{equation}\label{eq:setprob}
  \Pset \equiv \left\{\left[x_1, \ldots, x_d \right]: x_i \geq x_{i+1} \geq 0 \ \mbox{and} \ \sum_{i=1}^d x_i=1  \right\}.
\end{equation}
Geometrically, this set is a convex polytope embedded in the
$d-1$-probability simplex.

Let us now introduce the notion of majorization between probability
vectors (see, e.g.~\cite{MarshallBook}).

\begin{definition}
\label{def:majorization}
For given $x,y \in \Pset$, it is said that
$x$ majorizes $y$,
denoted as $x \succeq y$, if and only if,
\begin{equation}
\label{eq:partialsums}
\sum_{i=1}^k x_{i} \geq \sum_{i=1}^k y_{i} \ \forall \, k=1, \ldots, d-1.
\end{equation}
\end{definition}

Notice that $\sum_{i=1}^d x_{k} = \sum_{i=1}^d y_{k}$ is trivially satisfied, because $x$
and $y$ are probability vectors (so we can discard this
condition from the definition of majorization).

The intuitive idea of majorization is that a probability
distribution majorizes another one, whenever the former is more
concentrated than the latter. In this sense, majorization provides a
quantification of the notion of non-uniformity. To fix ideas, let us
observe that any probability vector $x \in \Pset$ trivially
satisfies the majorization relations: $e_d \equiv [1, 0, \dots, 0]
\maj x \maj \left[\frac{1}{\rank x}, \ldots, \frac{1}{\rank x}, 0,
\ldots,0 \right]  \maj \left[\frac{1}{d}, \ldots, \frac{1}{d}
\right] \equiv u_d$, where $\rank x$ is the number of positives
entries of $x$, and $e_d$ and $u_d$ are the extreme $d$-dimensional
probability vectors in the sense of maximum non-uniformity ($e_d$)
and minimum non-uniformity ($u_d$, i.e. the uniform probability
vector), respectively.
Let us remark that there are several equivalent definitions of
majorization that connect it with the notions of double stochastic
matrices, Schur-concave functions and entropies, among others (see
\eg \cite{MarshallBook}).

Here we are interested in the order-theoretic properties of
majorization. Indeed, it can be shown that the set $\Pset$ together
with the majorization relation is a \textit{partially ordered set}
(POSET, see \eg~\cite{DaveyBook} for an introduction to order theory).
This means that that, for every $x,y,z \in \Pset$ one has
\begin{enumerate}[(i)]
  \item reflexivity: $x \maj x$,
  \item antisymmetry: $x \maj y$ and $y \maj x$, then $x=y$, and
  \item transitivity: $x \maj y$ and $y \maj z$, then $x \maj  z$.
\end{enumerate}
Notice that if one leaves the
constraint that the entries of the probability vectors are sorted in
non-increasing order, then condition (ii) is not valid in general.
Instead of this, a weaker version holds, where $x$ and $y$ differ only
by a permutation of its entries. In such case, majorization gives a
preorder because condition (i) and (iii) remain valid.

In general, majorization does not yields a total order for
probability vectors belonging to $\Pset$. This is because there
exist $x,y \in \Pset$ such that $x \nmaj y$ and $y \nmaj x$ for any
$d >2$. In this situation, we say that the probability vectors are
incomparable. For instance, it is straightforward to check that
$x=[0.6, 0.16, 0.16, 0.08]$ and $y=[0.5, 0.3, 0.1, 0.1]$ are
incomparable.

\begin{figure}[h]
  \centering
  \includegraphics[width=.7\textwidth]{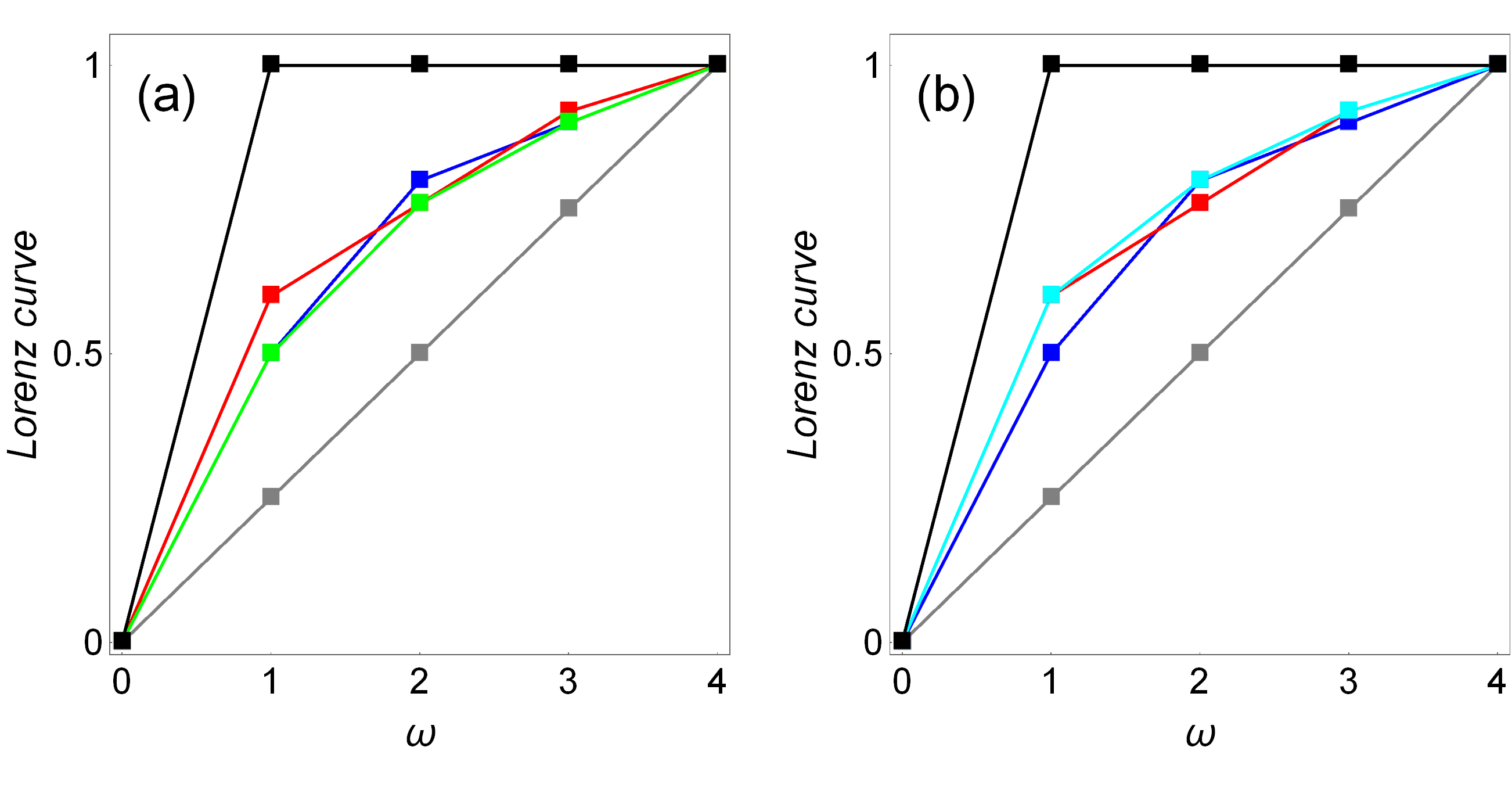}\\
  \caption{Lorenz curves of of $e_4$ (black), $u_4$ (gray), $x=[0.6, 0.16, 0.16, 0.08]$ (red) and $y=[0.5, 0.3, 0.1, 0.1]$ (blue).
  (a) Among all Lorenz curves below the ones of $x$ and $y$, there exists the greatest Lorenz curve that corresponds to the probability vector $x \wedge y = [0.5, 0.26, 0.14, 0.1]$  (green).
  (b) Among all Lorenz curves above the ones of $x$ and $y$, there exists the lowest Lorenz curve that corresponds to the probability vector $x \vee y = [0.6, 0.2, 0.12, 0.08]$  (cyan).
  }\label{fig:lorenzcurve}
\end{figure}

There is a visual way to address majorization that consists in
appealing to the notion of Lorenz curve \cite{Lorenz1905}. More
precisely, for a given $x \in \Pset$ one introduces the set of
points $\left\{(k,\sum_{i=1}^k x_i)\right\}_{k=0}^d$ (with the
convention $(0,0)$ for $k=0$). Then, the Lorenz curve of $x$, say
$L_x(\omega)$ with $\omega \in  [0,d]$, is obtained by the linear
interpolation of these points. At the end, one obtains a
non-decreasing and concave polygonal curve from $(0,0)$ to $(d,1)$.
In this way, given two Lorenz curves of $x$ and $y$, if the Lorenz
curve of $x$ is greater (or equal) than the one of $y$, it implies that $x$
majorizes $y$, and vice versa. 
On the other hand, if two \emph{different} Lorenz curves intersect at least at one point
in the interval $(1,d)$, it means that $x$ and $y$ are incomparable.
See for example Fig.~\ref{fig:lorenzcurve}, where the Lorenz curve of $e_4, u_4,
x=[0.6, 0.16, 0.16, 0.08]$ and $y=[0.5, 0.3, 0.1, 0.1]$ are plotted.
It is clear that $e_4 \maj x \maj u_4$ and $e_4 \maj y\maj u_4$, but
$x \nmaj y$ and $y \nmaj x$. However, in such case, one can easily
realize that there are infinite Lorenz curves below the ones of $x$
and $y$, and among of all them, there is one which is the greatest
one. In the same vein, there are infinitely many Lorenz curves above
those of $x$ and $y$, and there is one which is the lowest one.

These intuitions can be formalized and allow to formulate a notion of infimum and supremum in the general case~\cite{Cicalese2002,Bapat1991,Bondar1994}.
Consequently, the definition of majorization lattice is introduced as follows:
\begin{definition}
The quadruple $\La = \langle \Pset, \maj, e_d,u_d \rangle $ defines a bounded lattice order structure, where $e_d = [1,0, \ldots,0]$ is the top element, $u_d = \left[\frac{1}{d}, \ldots, \frac{1}{d}\right]$ is the bottom element and for all $x,y \in \Pset$ the \textit{infimum} $x\wedge y$ and the \textit{supremum} $x\vee y$ are
expressed as in \cite{Cicalese2002} (or see below).
\end{definition}

Precisely, the components of
the infimum are given by iteration of the formula
\begin{equation}\label{eq:inf}
  (x \wedge y)_k = \min \left\{ \sum^k_{i=1} x_i, \sum^k_{i=1} y_i \right\} - \min \left\{ \sum^{k-1}_{i=1} x_i, \sum^{k-1}_{i=1} y_i \right\},
\end{equation}
for $k=1, \ldots, d$ and the convention that summations with the
upper index smaller than the lower index are equal to zero. For the
supremum, one has to proceed in two steps. First, one has to
calculate the probability vector, say $z$, with components given by
\begin{equation}\label{eq:zmax}
  z_k = \max \left\{ \sum^k_{i=1} x_i, \sum^k_{i=1} y_i \right\} - \max \left\{ \sum^{k-1}_{i=1} x_i, \sum^{k-1}_{i=1} y_i \right\}.
\end{equation}
In general, this vector does not belong to $\Pset$, because its
components are not in a decreasing order. If it is the case that $z
\in \Pset$, then $z = x \vee y$. Otherwise, one has to apply the
flatness process (see~\cite[Lemma 3]{Cicalese2002}) in order to get
the supremum, as follows. For a probability vector $w = [w_1,
\ldots, w_d]^t$, let $j$ be the smallest integer in $[2, d]$ such
that $w_j > w_{j-1}$ and let $k$ be the greatest integer in $[1,
j-1]$ such that
\begin{equation}
\label{eq:flat1}
w_{k-1} \geq \frac{\sum_{l=k}^j w_l }{j-k+1} = a,
\end{equation}
with $w_0 > 1$.
Then, a flatness probability vector $w'$ is given by
\begin{equation}
\label{eq:flat2}
w'_{l} = \left\{ \begin{array}{cc}
 a & \mbox{for} \ l=k,k+1, \ldots, j \\
 w_l & \mbox{otherwise}.
 \end{array}
 \right.
\end{equation}
Then, the supremum is obtained in no more than $d-1$ iterations, by
iteratively applying the above transformations with the input
probability vector $z$ given by \eqref{eq:zmax}, until one obtains a
probability vector in $\Pset$.

Let us consider a \emph{finite} set of probability vectors, that is,
$\P = \{ x^1, \ldots, x^N \}$ with $\ x^i \in \Pset$. By appealing
to the algebraic properties of the definition of lattice, it is
straightforward to show that the infimum and the supremum of $\P$
always exist, and are given by $\bigwedge \P = x^1 \wedge x^2 \wedge
\ldots \wedge x^N$ and $\bigvee \P = x^1 \vee x^2 \vee \ldots \vee
x^N$. However, if one considers an arbitrary set of probability
vectors (which could be infinite), the lattice properties are not
strong enough to guarantee the existence of infimum and supremum. If
the infimum and supremum exist for arbitrary families, the lattice
is said to be \emph{complete}. It has been shown that the majorization lattice is indeed complete
\cite{Bapat1991,Bondar1994}.
For the sake of completeness, we reproduce the demonstration here and extend it in the following
sense: we provide an explicit algorithm for computing the supremum.
\begin{proposition}
\label{lemma:completness}
Let $\P$ an arbitrary set of probability vectors such that $\P
\subseteq \Pset$. Then, there exist the infimum $x^{\inf} \equiv
\bigwedge \P$ and the supremum $x^{\sup} \equiv \bigvee \P$ of $\P$.

\noindent In addition, the components of the $x^{\inf}$ are given by
\begin{equation}\label{eq:infuncountableset}
  x^{\inf}_k = \inf \left\{ \S_k \right\} - \inf \left\{\S_{k-1} \right\},
\end{equation}
where $\S_k = \{S_k(x): x \in \P \}$ with $S_k(x) \equiv
\sum^k_{i=1} x_i$ for $k \in \{1, \ldots, d\}$ and $S_0(x) \equiv
0$.

\noindent On the other hand, to obtain the components of the $x^{\sup}$, we have first to define the probability vector with components given by
\begin{equation}\label{eq:supuncountableset1}
  \bar{x}_k = \sup \left\{ \S_{k} \right\} - \sup \left\{\S_{k-1} \right\}.
\end{equation}
Then, we compute the upper envelope of the polygonal given by the linear interpolation of the points $\{(k, S_k(\bar{x})) \}_{k=0}^d$, say $\bar{L}(\omega)$, by using the algorithm~\ref{alg:upperenv}.
Finally, the components of the supremum are given by:
\begin{equation}\label{eq:supuncountableset}
  x^{\sup}_k =  \bar{L}(k) - \bar{L}(k-1).
\end{equation}
\end{proposition}

The proof of Proposition~\ref{lemma:completness} is given in~\ref{app:proofcompletness}.
Clearly, when the set is given by two probability vectors in
$\Pset$, that is $\P = \{x,y\}$, the calculus of infimum and
supremum of the Proposition~\ref{lemma:completness} reduces to the
procedure given in Ref.~\cite{Cicalese2002} (see
Eqs.~\eqref{eq:inf}--\eqref{eq:flat2}).

\subsection*{Infimum and supremum over convex polytopes}

\begin{figure}[h]
  \centering
  \includegraphics[width=.7\textwidth]{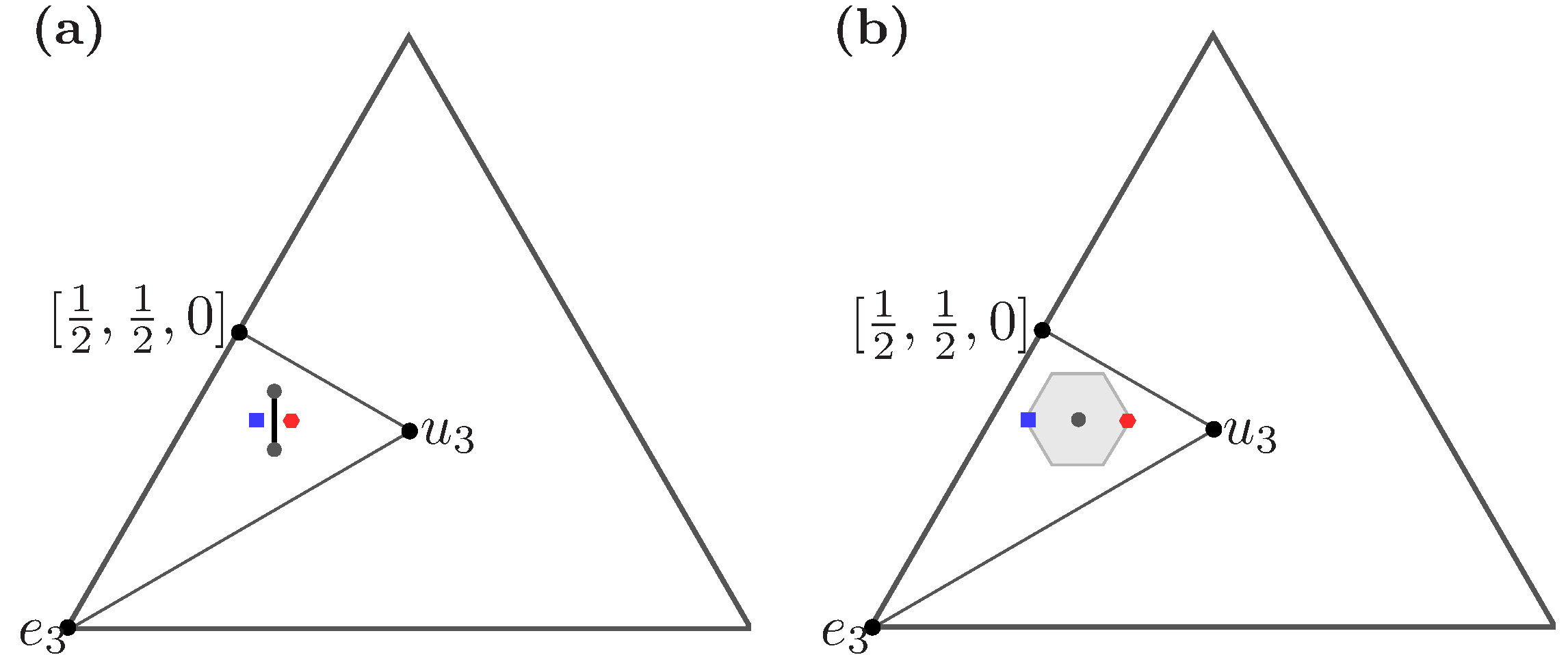}\\
  \caption{Infimum and supremum  of convex polytopes in $\Delta^\downarrow_3$ (region formed by the convex hull of $e_3$, $u_3$ and $\left[\frac{1}{2},\frac{1}{2},0 \right]$) for (a) $\P= \left\{x \in \Delta^\downarrow_3: x = p [0.5, 0.4, 0.1] + (1-p) [0.55, 0.3, 0.15] \, \mbox{with} \, p \in [0,1]\right\}$ (black line), where $\bigwedge \P = [0.5, 0.35, 0.15]$ (red hexagon) and $\bigvee \P = [0.55, 0.35, 0.1]$ (blue square); and (b) $\P = B_{0.15}([0.525, 0.35, 0.125])$ (light gray region), where $\bigwedge \P = [0.45, 0.35, 0.2]$ (red hexagon) and $\bigvee \P = [0.6, 0.35, 0.05]$ (blue square).
  }\label{fig:bola}
\end{figure}

Let us illustrate the meaning and relevance of the infimum and
supremum discussed above with an interesting example. First, let us
note that if $\P \subseteq \Pset$ is a convex polytope, then the
corresponding infimum and supremum can be computed as the infimum
and supremum of the set of vertices, $\mathrm{vert}(\P)$.
\begin{lemma}
\label{lemma:convexplytope}
Let $\P$ be a convex polytope contained in $\Pset$, and
$\mathrm{vert}(\P)$ the set of vertices, $\mathrm{vert}(\P) =
\{v^n\}^N_{n=1}$. Then, the infimum $x^{\inf} \equiv \bigwedge \P$
and the supremum $x^{\sup} \equiv \bigvee \P$ of $\P$ are given by
the infimum and supremum elements of $\mathrm{vert}(\P)$, namely
\begin{equation}
  x^{\inf} = \bigwedge \{ v^n \}^N_{n=1} \quad \mbox{and} \quad
  x^{\sup} = \bigvee \{ v^n\}^N_{n=1}.
\end{equation}
\end{lemma}
The proof of Lemma~\ref{lemma:convexplytope} is given in \ref{app:lemma1}.
Notice that, although the problem is reduced to the calculation of
the infimum and supremum among the extreme points of the convex
polytope, $x^{\inf}$ and $x^{\sup}$ do not necessarily belong to it (see \eg, Fig.~\ref{fig:bola}.(a)).
However, we will see an interesting example where the infimum and
supremum do belong to the given convex polytope (see \eg, Fig.~\ref{fig:bola}.(b)).

Let us consider the $\ell_1$--norm $\epsilon$--ball centered in $x^0
\in \Pset$ intersected with $\Pset$, that is, $B_\epsilon(x^0) =\{ x'
\in \Pset: || x' - x^0||_1 \leq \epsilon \}$, where $||x||_1 \equiv
\sum^d_{i=1} |x_i|$ denotes the $\ell_1$--norm of a probability
vector. Let us first note that $\{ x' \in \mathbb{R}^d: || x' -
x^0||_1 \leq \epsilon \}$ is a convex polytope (see
Ref.~\cite{Hanson2018}). Then, $B_\epsilon(x^0)$ is also a convex
polytope, because it is the intersection of that convex polytope
with $\Pset$. Therefore, by applying
Lemma~\ref{lemma:convexplytope},  $\bigwedge B_\epsilon(x^0)$ and
$\bigvee B_\epsilon(x^0)$ reduces to finding the infimum and
supremum of the vertices of $B_\epsilon(x^0)$.

Interestingly enough, the lattice-theoretic property of majorization can be posed in
strong connection with the notion of approximate majorization~\cite{Hanson2018,Horodecki2018}, which has recently
found application in quantum thermodynamics~\cite{vanderMeer2017}. More precisely, the
steepest $\epsilon$-approximation, $\bar{x}^{0(\epsilon)} \in
B_\epsilon(x^0)$, and the flattest $\epsilon$-approximation,
$\underline{x}^{0(\epsilon)} \in B_\epsilon(x^0)$, of $x^0$ given
in~\cite{Hanson2018,Horodecki2018} satisfy  that,
$\bar{x}^{0(\epsilon)} \maj x \maj \underline{x}^{0(\epsilon)}$ for
all $x \in  B_\epsilon(x^0)$. Using the definition of infimum and
supremum of a given family, it follows that
$\bar{x}^{0(\epsilon)}=x^{\sup}$ and
$\underline{x}^{0(\epsilon)}=x^{\inf}$, although the algorithms to
obtain them are different to the ones presented here. Thus, we see
that the notion of approximate majorization is in strong connection
with the property of completeness of the majorization lattice.
Furthermore, we have shown that it can be reduced to the application
of the algorithm of infimum and supremum to the set of vertices of
$B_\epsilon(x^0)$.

\section{Optimal common resource}

Now, we are ready to apply the Proposition~\ref{lemma:completness} to the problem of finding the optimal common resource in QRTs based on majorization.

In the first place, we have to distinguish between two possible cases of QRTs based on majorization.
We call direct majorization-based QRTs to those QRTs such that $\rho \free \sigma$ iff $x(\rho) \maj x(\sigma)$, whereas we call reversed majorization-based QRTs to those that  reverse the majorization relation (that is, $\rho \free \sigma$ iff $x(\sigma) \maj x(\rho)$).
Notice that purity is of the former type, whereas entanglement and coherence are of the latter one (see Table~\ref{tab:QRT maj}).
For such QRTs, let us remark that $\rho^{\ocr}$ is an optimal common resource if $\rho^{\ocr} \free \sigma \ \forall \sigma
\in \T$, and for any other $\rho$ satisfying $\rho \free \sigma \ \forall \sigma \in \T$ one has $\rho \free \rho^\ocr$.
Let us observe that all states $\rho$ such that $\rho^{\ocr} \bifree \rho$ are equivalent in the sense that all of them are optimal common resources.

For a given set of target resources $\T$, let us consider its corresponding set of probability vectors $\P$, which depends on the majorization-based QRT that one is dealing with.
We show now that the problem of finding an optimal common resource within a QRT based on majorization, can be reduced to an application of the
completeness of the majorization lattice.
Indeed, by directly applying Proposition~\ref{lemma:completness}, one finds that an optimal common resource of $\T$ for direct majorization-based QRTs can be obtained from the supremum of the corresponding set of probability vectors $\P$.
On the other hand, for reversed majorization-based QRTs, it can be obtained from the infimum of the corresponding set of probability vectors $\P$.

In this way, the completeness of the majorization lattice is of the essence in dealing with the optimal common resources in QRTs based on majorization.
As we have already stressed in the Introduction, this is a twofold extension of the proposal of Ref.~\cite{Guo2018}.

\subsection*{Optimal common resource within the resource theory of quantum coherence}

In the following, we illustrate how to obtain an optimal common resource within the resource theory of quantum coherence introduced in Ref.~\cite{Baumgratz2014}.

Deterministic transformations between pure sates by means of incoherent operations (free operations) have been addressed in several works~\cite{Du2015,Chitambar2016,Du2017,Zhu2017}.
In particular, we consider two pure sates $\ket{\psi} = \sum_{i=1}^d \psi_i \ket{i}$ and $\ket{\phi} = \sum_{i=1}^d \phi_i \ket{i}$, where
$\{\ket{i}\}_{i=1}^{d}$ is a fixed orthonormal basis (the incoherent basis) of a $d$-dimensional Hilbert space. The coefficients $\{\psi_i\}$ and  $\{\phi_i\}$ are complex numbers in general, satisfying $\sum_{i=1}^d |\psi_i|^2 = \sum_{i=1}^d |\phi_i|^2 =1$.
Let $x(\psi)$ and $x(\phi)$ be the probability vectors in $\Pset$ associated to these pure states, that is, $x_i(\psi) = |\psi_{[i]}|^2$ and $x_i(\phi) = |\phi_{[i]}|^2$, where $|\psi_{[i+1]}| \geq |\psi_{[i]}|$ and $|\phi_{[i+1]}| \geq |\phi_{[i]}|$ for all $i$.
It has be shown that $\ket{\psi}$ can be converted into $\ket{\phi}$ by means of incoherent operations (IO), denoted as $\ket{\psi} \underset{\mathrm{IO}}{\rightarrow}\ket{\phi}$, if and only if $x(\phi)  \maj x(\psi)$ (see \eg Ref.~\cite{Streltsov2017} and references therein).
This result can be seen as the analog of the celebrated Nielsen's theorem~\cite{Nielsen1999} for quantum coherence.

Let us recall that $\sum_{i=1}^d \frac{1}{\sqrt{d}} \ket{i}$ is a maximally coherent state, since it can be converted into any other state by means of  incoherent operations~\cite{Baumgratz2014}. We are going to discuss two cases in which the optimal common resource is not a maximally coherent one.

As a first example, if we consider a subset of pure states given by $\T= \left\{\ket{\phi} \in \mathbb{C}^d: |\phi_{[1]}| \geq \alpha \right\}$ with $1/\sqrt{d}<\alpha \leq 1$, to find an optimal common resource of $\T$ we have to calculate the infimum of the set $\P = \{x \in \Pset: x_1 \geq \alpha^2\}$. It can be shown that $\bigwedge \P = \left[\alpha^2, \frac{1-\alpha^2}{d-1}, \ldots, \frac{1-\alpha^2}{d-1}\right]$, so that an optimal common resource has the form $\ket{\psi^\ocr} = \alpha \ket{1} + \sum_{i=2}^d \sqrt{\frac{1-\alpha^2}{d-1}} \ket{i}$. Clearly this optimal common resource is not a maximally coherent state.

As a second example, motivated by the study of coherence of quantum superpositions~\cite{Yue2017}, let us consider a more subtle target set formed by superpositions of given two orthogonal states.
More precisely, let $\T= \left\{\ket{\phi} \in \mathbb{C}^d: \ket{\phi} = \alpha \ket{\mu} + \beta \ket{\nu} \right\}$, where $\ket{\mu}= \sum_{i=1}^{d_1} \frac{1}{\sqrt{d_1}} \ket{i}$,  $\ket{\nu}= \sum_{i=d_1+1}^{d} \frac{1}{\sqrt{d-d_1}} \ket{i}$ and  $\alpha^2+\beta^2 = 1$ (with  $\alpha,\beta\in\mathbb{R}$  for simplicity). If we do not impose any other restriction over $\alpha$, then $\T$ contains the maximally coherent state, which is trivially the optimal common resource. In order to exclude that possibility, let us consider that $\alpha^2\neq d_1/d$. In particular, let us suppose that $\alpha^2>d_1/d$, so that there is $\alpha^2_{\min}$ such that $\alpha^2_{\min}\leq\alpha^2\leq1$, with $\alpha^2_{\min}$ strictly greater than $d_1/d$ (the other case, with $\alpha^2<d_1/d$, is completely analogous). The corresponding set of probability vectors is
\begin{equation*}
\P = \bigg\{x \in \Pset: x = \Big[\overbrace{\frac{\alpha^2}{d_1}, \ldots,\frac{\alpha^2}{d_1}}^{d_1},\frac{1-\alpha^2}{d-d_1},\ldots,\frac{1-\alpha^2}{d-d_1} \Big]  \bigg\},
\end{equation*}
and the infimum of $\Pset$ is shown to be
\begin{equation*}
\bigwedge \P = \Big[\overbrace{\frac{\alpha^2_{\min}}{d_1}, \ldots,\frac{\alpha^2_{\min}}{d_1}}^{d_1},\frac{1-\alpha^2_{\min}}{d-d_1},\ldots,\frac{1-\alpha^2_{\min}}{d-d_1} \Big].
\end{equation*}
Therefore, an optimal common resource of $\T$ is of the form
\begin{equation*}
\ket{\psi^\ocr} = \sum_{i=1}^{d_1} \sqrt{\frac{\alpha^2_{\min}}{d_1}} \ket{i} + \sum_{i=d_1+1}^d \sqrt{\frac{1-\alpha^2_{\min}}{d-d_1}} \ket{i}.
\end{equation*}
Notice that in this example $\ket{\psi^\ocr} \notin \T$.

\section{Concluding remarks}

In this paper we gave a solution for the problem of finding an optimal common resource for an arbitrary family of target states of
a given a QRT based on majorization like entanglement, coherence or purity (see Table~\ref{tab:QRT maj}).
Our method relies on the completeness properties of the majorization lattice.
We provided concrete algorithms for computing the infimum and supremum of an arbitrary family of states (Proposition~\ref{lemma:completness}). 
Our contribution improves previous works (e.g. \cite{Bapat1991,Bondar1994,Cicalese2002,Guo2018}), in the sense that
our algorithm works for target sets of arbitrary cardinality (i.e., we provide an expression for the supremum for
possibly non-denumerable families of states).
Also, for convex polytopes, we include a study of the relationship between
the infimum and supremum, and their extreme points (Lemma~\ref{lemma:convexplytope}).

In addition, we showed that the notion of approximate majorization is in strong connection with the property of completeness of the
majorization lattice~\cite{Hanson2018,Horodecki2018}.
Indeed, the flattest and steepest approximations are nothing more than the infimum and supremum of the corresponding set, respectively, and
they can be calculated only from their vertices.

Finally, the fact that completeness of the majorization lattice is of the essence in dealing with the optimal common resources is illustrated with some examples within the resource theory of quantum coherence~\cite{Baumgratz2014}.

\section*{Acknowledgement}
This work has been partially supported by CONICET (Argentina) and by Fondazione di Sardegna within the project ``Strategies and Technologies for Scientific Education and Dissemination'', cup: F71I17000330002.

\appendix

\section*{Appendix}
\setcounter{section}{1}

\subsection{Proof of Proposition~\ref{lemma:completness}}
\label{app:proofcompletness}

For the sake of completeness, we show here that the majorization lattice is complete.
We stress that this has been proved in previous works \cite{Bapat1991,Bondar1994}.
However, we give an alternative proof that relies heavily on the more geometric construction provided by the Lorenz curves.
This allow us to provide an explicit algorithm for the computation not only of the infimum but also of the supremum (Proposition~\ref{lemma:completness}).

Let us first introduce some notations and definitions. Let us define
the partial sum of the first $k$ components of a given vector $x$ as
$S_k(x) \equiv \sum^k_{i=1} x_i$ with the convention $S_0(x) \equiv
0$. Now, let us consider the set formed by all partial sums up to
$k$ that come from probability vectors in $\P$, that is, $\S_k=
\{S_k(x)\,:\,x \in \P  \}$ and its infimum $\lbar{S}_k \equiv \inf
\S_k$ and supremum $\bar{S}_k \equiv \sup \S_k$. Notice that, for
each $k=0, \ldots d$, both $\lbar{S}_k$ and $\bar{S}_k$ exist, since
each $\S_k$ is a set of real numbers bounded from below by
$\frac{k}{d}$ and above by $1$. Finally, let us consider the
probability vectors $\lbar{x}=[\lbar{S}_1, \lbar{S}_2 - \lbar{S}_1,
\ldots,\lbar{S}_i - \lbar{S}_{i-1} ,\ldots \lbar{S}_{d} -
\lbar{S}_{d-1}]$ and $\bar{x}=[\bar{S}_1, \bar{S}_2 - \bar{S}_1,
\ldots,\bar{S}_i - \bar{S}_{i-1} ,\ldots \bar{S}_{d} -
\bar{S}_{d-1}]$. Let us prove that from these probability vectors
one can obtain the infimum and the supremum, respectively.

\hfill

\subsubsection{Infimum}

Let us now prove that $\lbar{x}=x^{\inf}$. To prove that, we appeal
to the description of majorization in terms of Lorenz curves. First
we show that the curve $L_{\lbar{x}}(\omega)$ with $\omega \in
[0,d]$, formed by the linear interpolation of the points $\left\{(k,
\lbar{S}_k)\right\}_{k=0}^d$ (notice that $\lbar{S}_0 = 0$ and
$\lbar{S}_d =1$) is a Lorenz curve. This is equivalent to prove that
$\lbar{x} \in \Pset$. We proceed in two steps: (a)
$L_{\lbar{x}}(\omega)$ is non-decreasing \ie, $L_{\lbar{x}}(k) \leq
L_{\lbar{x}}(k+1)$ for all $k \in \{0, \ldots, d-1\}$ (b)
$L_{\lbar{x}}(\omega)$ is concave \ie, $L_{\lbar{x}}(k) \geq
\frac{1}{2}\left(L_{\lbar{x}}(k-1)+L_{\lbar{x}}(k+1) \right)$ for
all $k \in \{1, \ldots, d-1\}$. The proofs of both points are given
by \emph{reductio ad absurdum}.

Let us proceed with the proof of (a) $L_{\lbar{x}}(k) \leq
L_{\lbar{x}}(k+1)$ for all $k \in \{0, \ldots, d-1\}$. Let us assume
that there exists  $k'$ such that $L_{\lbar{x}}(k')>
L_{\lbar{x}}(k'+1)$. By construction, there exists a sequence, say
$\{ L_{x^i}(k'+1)  \}_{i \in \mathbb{N}}$ with $x^{i} \in \P $, of
elements of $\S_{k'+1}$, that converges to $\lbar{S}_{k'+1}=
L_{\lbar{x}}(k'+1)$. Let us choose $i'$ big enough such that
$L_{x^i}(k'+1) - L_{\lbar{x}}(k'+1) \leq
\frac{1}{2}\left(L_{\lbar{x}}(k') - L_{\lbar{x}}(k'+1) \right)$ for
all $i \geq i'$. Let us pick one of them, say $i_0$. On the other hand,
by definition of $\lbar{x}$, one has $L_{x^{i_0}}(k') \geq L_{\lbar{x}}(k')$.
Finally, one has $L_{x^{i_0}}(k') \geq L_{\lbar{x}}(k') >
L_{\lbar{x}}(k'+1) +\frac{1}{2}\left(L_{\lbar{x}}(k') -
L_{\lbar{x}}(k'+1) \right) \geq L_{x^{i_0}}(k'+1)$. But this is in
contradiction with the fact that $L_{x^{i_0}}(k) \leq L_{x^{i_0}}(k+1)$
for all $k \in \{0, \ldots, d-1\}$, which is true by definition of
Lorenz curve. Then, (a) holds.

Now, we proceed with the proof of (b): $L_{\lbar{x}}(k) \geq
\frac{1}{2}\left(L_{\lbar{x}}(k-1)+L_{\lbar{x}}(k+1) \right)$ for
all $k \in \{1, \ldots, d-1\}$ Assume that there exists $k'$ such
that $L_{\lbar{x}}(k') <
\frac{1}{2}\left(L_{\lbar{x}}(k'-1)+L_{\lbar{x}}(k'+1) \right)$. By
construction, there exists a sequence, say $\{ L_{x^i}(k')  \}_{i
\in \mathbb{N}}$ with $x^{i} \in \P $, of elements of $\S_{k'}$,
that converges to $\lbar{S}_{k'}= L_{\lbar{x}}(k')$. Let us choose
$i'$ big enough such that $L_{x^i}(k') <
\frac{1}{2}\left(L_{\lbar{x}}(k'-1)+L_{\lbar{x}}(k'+1)\right)$ for
all $i \geq i'$. Let us pick one of them, say $i_0$. By definition of $\lbar{x}$,
$L_{x^{i_0}}(k'-1) \geq L_{\lbar{x}}(k'-1)$ and $L_{x^{i_0}}(k'+1)
\geq L_{\lbar{x}}(k'+1)$. This implies that $L_{x^{i_0}}(k') <
\frac{1}{2}\left(L_{\lbar{x}}(k'-1) +  L_{\lbar{x}}(k'+1) \right)
\leq \frac{1}{2}\left(L_{x^{i_0}}(k'-1) + L_{x^{i_0}}(k'+1)
\right)$. But this is in contradiction with the fact $L_{x^{i_0}}(k)
\geq \frac{1}{2}\left(L_{x^{i_0}}(k-1) + L_{x^{i_0}}(k+1) \right)$
for all $k \in \{1, \ldots, d-1\}$, which is true by definition of
Lorenz curve. Then,  (b) holds.

Up to now, we have proved that $L_{\lbar{x}}(\omega)$ is a Lorenz
curve that, by construction, satisfies $L_{\lbar{x}}(\omega) \leq
L_{x}(\omega) \ \forall \omega \in [0,d]$ and $\forall x \in \P$. In
other words, we obtain that $\lbar{x} \in \Pset$ and $x \maj
\lbar{x} \ \forall x \in \P$. It remains to be proved that for any
$x' \in \Pset$ such that $x \maj x' \ \forall x \in \P$, one has
$\lbar{x} \maj x'$. In order to do this, we appeal again to the
\emph{reductio ad absurdum} and the notion of Lorenz curve. Let us
assume that there exist $x'$ such that $x \maj x' \ \forall x \in
\P$, but $\lbar{x} \not\maj x'$.
This happens if at least one partial sum of $x'$ is greater than the
one of the $\lbar{x}$, say the $k'$ partial sum. In other words,
$L_{x'}(k') > L_{\lbar{x}}(k')$. Choose again a sequence $\{
L_{x^i}(k')\}_{i\in\mathbb{N}}$ of elements of $\S_{k'}$ that
converges to $\lbar{S}_{k'}$. Choose $i'$ big enough such that
$L_{x^i}(k') < L_{\lbar{x}}(k') + \frac{1}{2}\left( L_{x'}(k') -
L_{\lbar{x}}(k')\right)$ for all $i \geq i'$. Let us pick one of
them, say $i_0$, so $L_{x^{i_0}}(k') < L_{\lbar{x}}(k') +
\frac{1}{2}\left( L_{x'}(k') - L_{\lbar{x}}(k')\right)$. But, by
hypothesis, one has $L_{x'}(k) \leq L_{x^{i_0}}(k)$ for all $k \in
\{0, \ldots, d\}$, which is in contradiction with the previous
inequality. Thus, there does not exist such $x'$. Therefore,
$\lbar{x}=x^{\inf}$.

\subsubsection{Supremum}

Notice that, according to lattice theory, the arbitrary supremum can
be expressed in terms of the arbitrary infimum, and vice versa
\cite{BurrisBook}. This means that our proof of the existence of the
infimum for an arbitrary set $\P$ of probability vectors (whose
components are arranged in non-increasing order), automatically implies
the existence of its supremum $x^{\sup} = \bigwedge \{x' \in\Pset: x' \maj x \ \forall x \in \P \}$. With this observation we finish our proof that
the majorization lattice is complete. Notice that the mere proof of
the existence of a supremum, does not guarantee the existence of an
algorithm to compute it. Thus, in the sequel, we focus our efforts
in providing such an algorithm.

Consider the polygonal curve $L_{\bar{x}}(\omega)$, with $\omega \in
[0,d]$, formed by the linear interpolation of the points $\left\{(k,
\bar{S}_k)\right\}_{k=0}^d$ (notice that $\bar{S}_0 = 0$ and
$\bar{S}_d =1$). By construction, $L_{\bar{x}}(\omega)$ is
non-decreasing and satisfies that $L_{\bar{x}}(\omega) \geq
L_{x}(\omega), \, \forall \omega \in [0,d]$ and $\forall x \in \P$.
But, alike $L_{\lbar{x}}(\omega)$, $L_{\bar{x}}(\omega)$ is not
necessarily a Lorenz curve. Thus, it cannot be used to construct the
(ordered) probability vector associated to the supremum of the given
family. Instead, let us show that the upper envelope of
$L_{\bar{x}}(\omega)$, that is, $\bar{L}(\omega) \equiv \inf
\{g(\omega): g \mbox{ is concave and } g(\omega) \geq
L_{\bar{x}}(\omega) \, \forall \omega \in [0,d] \}$ (see \eg,
\cite[Def. 4.1.6]{BratteliBook}), is indeed the Lorenz curve
associated to the supremum: $L_{x^{\sup}}(\omega) =
\bar{L}(\omega)$. In this way, from the upper envelope
$\bar{L}(\omega)$, one obtains the supremum as $x^{\sup}
=[\bar{L}(1), \bar{L}(2) - \bar{L}(1), \ldots,\bar{L}(i) -
\bar{L}(i-1) ,\ldots, \bar{L}(d) - \bar{L}(d-1)]$. Thus, we have
to prove that: (a) $\bar{L}(\omega)$ is a Lorenz curve, and (b) if
$x' \in \Pset$ and $x' \maj x \ \forall x \in \P$, then $x' \maj
x^{\sup}$.

Our method to obtain the supremum $x^{\sup}$ has three steps: first,
we calculate $\bar{x}$; second, we compute the upper envelope of
$L_{\bar{x}}(\omega)$, $\bar{L}(\omega)$; third, we compute the
elements of $x^{\sup}$ as the components of the probability vector
associated to the Lorenz curve $\bar{L}(\omega)$. The first and last
steps are straightforward. We also provide the
algorithm~\ref{alg:upperenv} to find the upper envelope of a
polygonal curve with coordinates $\{(k,S_k(x))\}_{k=0}^d$.

\begin{algorithm}
    \caption{Upper envelope}
    \label{alg:upperenv}
    \begin{algorithmic}
        \State \textbf{input:} $x\in\mathbb{R}^d$
        \State \textbf{output:} coordinates of the upper envelope of the polygonal curve joining $\{(k,S_k(x))\}_{k=0}^d$.
        \smallskip
        \Procedure{UpperEnv}{$x$} 
        \State $\K \leftarrow \{0\}$ \Comment{Stores the `critical points' of $x$}
        \State $i \leftarrow 0$
            \While{$i < \mathrm{length}(x)$}
                \State $m \leftarrow \{0\}$ \Comment{Stores slope values}
                \For{$j = i\!+\!1\,\ldots\,\mathrm{length}(x)$}
                    \State $m \leftarrow \mathrm{append}\left\{m,\frac{S_j(x)-S_i(x)}{j-i}\right\}$
                \EndFor
                \State $k \leftarrow$ $\max$(position of $\max(m)$) \Comment{Finds position of the last maximum slope}
                \State $\K \leftarrow \mathrm{append}\{\K,k\}$
                \State $i \leftarrow k$ \Comment{Updates $i$}
            \EndWhile
            \State \textbf{return} $\{(k,S_{k}(x))\}_{k \in \K}$ \Comment{Coordinates of the upper envelope}
        \EndProcedure
    \end{algorithmic}
\end{algorithm}

Notice that for a given probability vector $\bar{x} \in
\mathbb{R}^d$, the output of the algorithm~\ref{alg:upperenv} is a
set of points $\{(k,S_{k}(\bar{x}))\}_{k \in \K}$. It is clear that
the linear interpolation of these points is a Lorenz curve, say
$L_{x^{\up}}(\omega)$, which has associated some probability vector
$x^\up \in \Pset$. Let us show that $L_{x^{\up}}(\omega)$ is equal
to the upper envelope of $L_{\bar{x}}(\omega)$. To see that, take
two consecutive indices, $k_i,k_{i+1}\in\K$. By construction,
$L_{\bar{x}}(\omega)=L_{x^{\up}}(\omega)$ for $\omega=k_i$ and
$\omega=k_{i+1}$. For $\omega\in[k_i,k_{i+1}]$,
$L_{x^{\up}}(\omega)$ is the linear interpolation and so one has two
possibilities: either $k_{i+1}=k_i+1$ and
$L_{\bar{x}}(\omega)=L_{x^{\up}}(\omega)$ for all
$\omega\in[k_i,k_{i+1}]$, or $k_{i+1}>k_i+1$ and
$L_{\bar{x}}(\omega)<L_{x^{\up}}(\omega)$ for some integer
$\omega\in(k_i,k_{i+1})$. In both cases, since the interpolation is
linear, there is no concave curve such that $L_x(\omega)\geq
L_{\bar{x}}(\omega)$ and $L_x(\omega)<L_{x^{\up}}(\omega)$ for all
$\omega\in(k_i,k_{i+1})$.
Since this is the case for any $k_i\in\K$, we necessarily obtain the upper envelope of the polygonal curve joining $\{(k,S_k(x))\}_{k=0}^d$.
Then, we have proved that $L_{x^{\up}}(\omega) = \bar{L}(\omega)$.
This last equality implies in turn that (a) $\bar{L}(\omega)$ is a Lorenz
curve. As a consequence, by construction of $\bar{L}(\omega)$, we
also have that $x^{\up} =[\bar{L}(1), \bar{L}(2) - \bar{L}(1),
\ldots,\bar{L}(i) - \bar{L}(i-1) ,\ldots, \bar{L}(d) -
\bar{L}(d-1)]$ satisfies that $x^{\up} \maj x \ \forall x \in \P$.
In addition, we have that $\nexists\, x' \in \Pset$ such that
$L_{x'}(\omega) \geq L_{\bar{x}}(\omega) \ \forall \omega \in [0,d]$
and $x^\up \maj x'$. Therefore, (b) holds and $x^{\up} = x^{\sup}$.

\subsection{Proof of Lemma~\ref{lemma:convexplytope}}
\label{app:lemma1}

We prove now that $x^{\inf} \equiv \bigwedge \P = \bigwedge
\mathrm{vert}(\P) \equiv v^{\inf}$ and $x^{\sup} \equiv \bigvee \P =
\bigvee \mathrm{vert}(\P) \equiv v^{\sup}$, that is to say that
infimum and supremum can be computed among the set of vertices of
the convex polytope.

Let $x$ be an arbitrary probability vector in $\P\subseteq\Pset$.
Since $\P$ is a convex polytope, $x$ can be written as a convex
combination of the vertices, $x=\sum_{n=1}^N{p_nv^n}$, with
$v^n\in\mathrm{vert}(P)$, $p_n\geq0$ and $\sum_{n=1}^Np_n=1$. For arbitrary $k$, the $k$-partial sum of $x$
gives
\begin{equation} \label{eq:proof_lemma1_1}
  S_k(x) = \sum_n p_n S_k(v^n) \geq S_k(v^{\inf}), \quad \forall k,\,\forall x\in\P,
\end{equation}
where we have used that, by definition, $v^n \maj v^{\inf}$,
$\forall v^n\in\mathrm{vert}(\P)$. On the other hand, since
$\mathrm{vert}(\P)\subseteq\P$ and given that
$x^{\inf}\equiv\bigwedge\P$, we know by definition of infimum that
$v^{\inf} \maj x^{\inf}$ must hold. Hence,
using~\eqref{eq:proof_lemma1_1},
\begin{equation}
  x \maj v^{\inf} \maj x^{\inf}, \quad \forall x\in\P.
\end{equation}
Therefore, by definition of infimum, one has $x^{\inf} = v^{\inf}$.

Analogously, for the supremum one obtains that
\begin{equation}
  x^{\sup} \maj v^{\sup} \maj x, \quad \forall x\in\P,
\end{equation}
and the desired result follows as before, by definition of supremum, $x^{\sup} = v^{\sup}$.



\end{document}